\documentclass[prl,twocolumn,showpacs,preprintnumbers,amsmath,amssymb]{revtex4}
\usepackage{bm}
\usepackage{amsmath}
\usepackage{dcolumn}
\usepackage[dvips]{graphicx}
\begin{document}
\bibliographystyle{apsrev}

\title{Finite nuclear size and Lamb shift of $p$-wave atomic states}

\author{A. I. Milstein}
\email[Email:]{A.I.Milstein@inp.nsk.su}
\affiliation{Budker
Institute of Nuclear Physics, 630090 Novosibirsk, Russia}
\author{O. P. Sushkov}
\email[Email:]{sushkov@phys.unsw.edu.au}
\affiliation{School of
Physics, University of New South Wales, Sydney 2052, Australia}
\author{I. S. Terekhov}
\email[Email:]{I.S.Terekhov@inp.nsk.su}
\affiliation{ Novosibirsk
University, 630090 Novosibirsk, Russia}

\date{\today}
\begin{abstract}
We consider corrections to the Lamb shift of $p$-wave atomic
states due to the finite nuclear size (FNS). In other words, these
are radiative corrections to the  atomic isotop shift related to
FNS. It is shown that the structure of the corrections is qualitatively different from
that for s-wave states.
The perturbation theory expansion for the relative correction for a $p_{1/2}$-state 
starts from $\alpha\ln(1/Z\alpha)$-term, while for $s_{1/2}$-states it starts
from $Z\alpha^2$ term. Here $\alpha$ is the fine
structure constant and $Z$ is the nuclear charge. In the present
work we calculate the $\alpha$-terms for $2p$-states, the result
for $2p_{1/2}$-state reads
$(8\alpha/9\pi)[\ln(1/(Z\alpha)^2)+0.710]$. Even more interesting are
$p_{3/2}$-states. In this case the ``correction''  is by several
orders of magnitude larger than the ``leading'' FNS  shift.

\end{abstract}
 \maketitle


Experimental and theoretical investigation of the radiative shift
(Lamb shift) of energy levels in heavy atoms  is an important way
to test Quantum Electrodynamics in presence of a strong external
electric field. One of the effects related to this problem is a
dependence of the Lamb shift on the finite nuclear size (FNS). One
can also look at this effect from another point of view. It is
well known that there is an isotop shift of atomic levels due to
the FNS. The corrections we are talking about are the radiative
corrections to the isotop shift.

The corrections for $1s$-, $2s$-, and $2p$-states have been
calculated numerically, exactly in $Z\alpha$, in
Refs.\cite{Blun92,CJS93,LPSY}. The self-energy and the vertex
corrections to the FNS  effect for any $s$-wave state have been
calculated analytically in order $\alpha(Z\alpha)$ in
Refs.\cite{Pach93,PG}. However, the structure of the higher order
$Z\alpha$ corrections and, in particular, their logarithmic
dependence on the nuclear size has not been understood even for
s-states. Our interest to FNS radiative corrections has been
stimulated by our work on the radiative corrections to atomic
parity nonconcervation \cite{Mil}. Technically the parity
nonconservation effect has some  common  features with that of the
FNS radiative correction: in both cases the effective size of the
perturbation source is much smaller than the Compton wavelength
$\lambda_C$. In  the paper \cite{Mil} we have elucidated the
structure of higher order in $Z\alpha$ FNS radiative corrections
for s-electrons, and have calculated analytically
$\alpha(Z\alpha)$ and $\alpha(Z\alpha)^2\ln(\lambda_C/r_0)$
self-energy  and vertex FNS relative radiative corrections. Here $r_0$ is
the nuclear radius. In the present work we calculate FNS radiative
corrections for p-wave electrons. We demonstrate that the
structure of the corrections for p-wave states is very much
different from  that for the s-wave states. Physically it happens
because of different infrared behavior.

\begin{figure}[h]
\centering
\includegraphics[height=140pt,keepaspectratio=true]{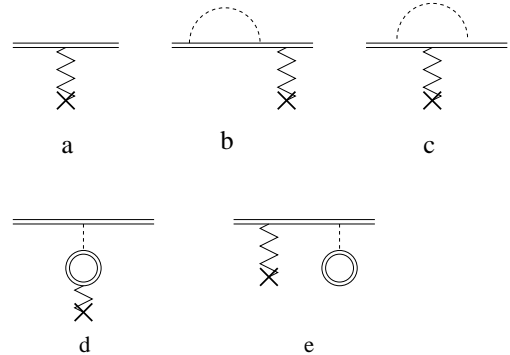}
\vspace{-10pt} \caption{\it The leading contribution to the FNS
effect is given  by diagram (a), and one loop radiative
corrections to the effect are given by diagrams (b-e). The double
line denotes the exact electron Green's function in the Coulomb
field of the nucleus, the cross denotes the nucleus, the zigzag
line denotes the FNS perturbation (\ref{dV}), and the dashed line
denotes the photon. } \label{Fig1} \vspace{-10pt}
\end{figure}

Due to the finite nuclear size, the electric potential $V(r)$ of
the nucleus is different from that for a pointlike nucleus. The
deviation is
\begin{equation}
\label{dV}
\delta V(r)=V(r)-\left(-\frac{Z\alpha}{r}\right)
\end{equation}
Throughout the paper we set $\hbar=c=1$. The diagram that describe
the FNS effect in the  leading order is shown in
Fig.\ref{Fig1}(a). The double line corresponds to the exact
electron wave function in the Coulomb field, and the zigzag line
with cross denotes the perturbation (\ref{dV}).

\noindent Diagrams Fig.\ref{Fig1}(b) and Fig.\ref{Fig1}(c)
correspond to the contributions of the electron self-energy
operator and the vertex operator, respectively. The diagram
Fig.\ref{Fig1}(d) describes a modification of $\delta V$ (see eq.
(\ref{dV})) due to the vacuum polarization, and  the diagram
Fig.\ref{Fig1}(e) corresponds to a modification of the electron
wave function due to the polarization of the vacuum by the Coulomb
field (Uehling potential).

\begin{figure}[h]
\centering
\vspace{30pt}
\includegraphics[height=200pt,keepaspectratio=true]{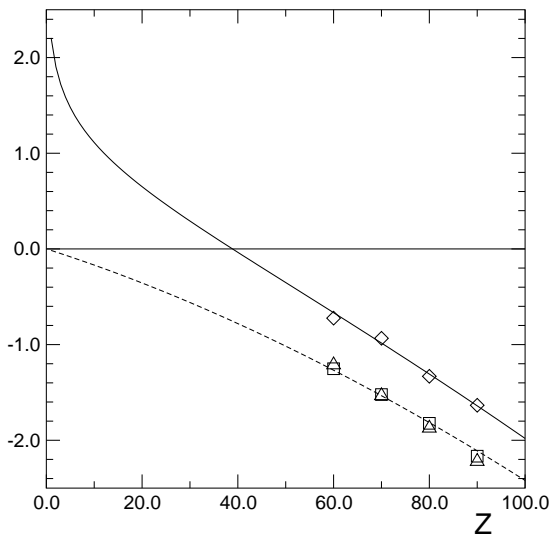}
\vspace{-0pt} \caption{\it Relative SEVFNS corrections (\%) for
$s_{1/2}$- and $p_{1/2}$-states. The dashed line shows the correction
$\Delta_s$ given by Eq. (\ref{s12}). Results of computations of
$\Delta_s$ for $1s$ and $2s$ states  \cite{CJS93} are shown by
squares and triangles, respectively. The solid line shows the
correction $\Delta_p$ given by Eq. (\ref{Dp}). Results of
computations of $\Delta_p$ for $2p_{1/2}$ state \cite{CJS93} are
shown by diamonds. } \label{Fig2} \vspace{5pt}
\end{figure}

Technically the most complicated  are the self-energy and the
vertex  FNS (SEVFNS) corrections given by diagrams in
Fig.\ref{Fig1}(b) and Fig.\ref{Fig1}(c). According to our previous
work \cite{Mil}, the  SEVFNS relative correction for an s-wave
state is of the form
 \begin{eqnarray}\label{s12}
\Delta_s&=&-\alpha\left[(Z\alpha)\left(\frac{23}{4}-4\ln2\right)\right.\nonumber\\
&+&\left.\frac{(Z\alpha)^2}{\pi}\left(\frac{15}{4}-\frac{\pi^2}{6}\right)
\ln(b\lambda_C/r_0) \right]\quad.
\end{eqnarray}
Here  $b=\exp(1/(2\gamma)-C-5/6)$, $C\approx 0.577$ is the Euler
constant, and $r_0$, as we already mentioned,  is the nuclear
radius. The total relative  SEVFNS correction (\ref{s12}) is the
ratio of the sum of diagrams Fig.\ref{Fig1}(b) and
Fig.\ref{Fig1}(c) divided by the diagram Fig.\ref{Fig1}(a). Value
of $\Delta_s$ is not proportional to the nuclear radius squared
because it is a relative quantity. Plot of $\Delta_s$ versus the
nuclear charge $Z$ is shown in Fig.\ref{Fig2} by the dashed line.
Results of computations of $\Delta_s$ for $1s$ and $2s$ states
\cite{CJS93} are shown by squares and triangles, respectively.
\noindent The $\alpha(Z\alpha)$ term in (\ref{s12}) comes from
distances $r\sim \lambda_C$, and the
$\alpha(Z\alpha)^2\ln(\lambda_C/r_0)$ term comes from distances
$r_0 \ll r \ll \lambda_C$. An important point is that there is no
contribution that comes from distances $r \gg \lambda_C$. Because
of this reason the correction $\Delta_s$ is exactly the same for
1s, 2s, 3s,... states \cite{com}. Why there is no contribution of
larger distances into $\Delta_s$? The reason is very simple. In
the leading order the correction $\Delta_s$ can be expressed in
terms of the forward electron-nucleus scattering amplitude
\cite{Mil}. There is a rigorous QED theorem that claims that there
is no an infrared divergence in the forward scattering amplitude,
see e.g. Ref. \cite{BLP}. Therefore, quantum fluctuations from
distances $r \gg \lambda_C$ cannot contribute to $\Delta_s$ (see
also Ref.\cite{LYE}). Let us look now at the p-wave SEVFNS
correction $\Delta_p$. From the point of view of the scattering
problem it corresponds to scattering at finite angle. The
finite-angle scattering amplitude is {\it always} infrared
divergent. Therefore, one {\it must} expect a contribution to
$\Delta_p$ from quantum fluctuations at distances $r \gg
\lambda_C$. This is the contribution we calculate in the present
work.

Formally we assume that $Z\alpha \ll 1$. Therefore, at distances
$r\sim \lambda_C/(Z\alpha) \gg \lambda_C$  dynamics of the
electron is described by usual  nonrelativistic Coulomb wave
functions. However, the nucleus radius is small, $r_0 \ll
Z\alpha\lambda_C \ll \lambda_C$. At so small distances, generally
speaking,  one must use  relativistic Dirac wave function even at
$Z=1$. The electron Dirac wave function at $r \ll
Z\alpha\lambda_C$ is of the form
\begin{equation}
\label{Dirac} \Psi({\bm r})= Nr^{\gamma-1}\begin{pmatrix}
{(\kappa-\gamma})\Omega\\
iZ\alpha\tilde{\Omega}
\end{pmatrix}
\quad ,
\end{equation}
where $\Omega$ and $\tilde{\Omega}=-({\bm \sigma}\cdot{\bm
n})\Omega$ are spherical spinors ; $\kappa=-1$ for
$s_{1/2}$-state, $\kappa=1$ for $p_{1/2}$-state, and $\kappa=-2$
for $p_{3/2}$-state ; $\gamma=\sqrt{\kappa^2-(Z\alpha)^2}$ ; and
$N$ is a constant known for each particular state, see Ref
\cite{BLP}. For $s_{1/2}$- and $p_{3/2}$-states the  upper
component of the Dirac spinor (\ref{Dirac}) is much larger than
the lower one. Hence, the upper component determines the FNS shift
of such a state. On the other hand, for $p_{1/2}$-state the lower
component and hence its contribution to the FNS shift is
dominating. A straightforward calculation gives the following
values for the FNS shifts of $2s$ and $2p$ states (diagram
Fig.\ref{Fig1}(a))
\begin{eqnarray}
\label{E0}
 \delta E^{(0)}(2s_{1/2})&=&\frac{1}{12}(Z\alpha)^4m^3<r^2>\, ,\nonumber\\
 \delta E^{(0)}(2p_{1/2})&=&\frac{1}{64}(Z\alpha)^6m^3<r^2>\, ,\nonumber\\
  \delta E^{(0)}(2p_{3/2})&=&\frac{1}{480}(Z\alpha)^6m^5<r^4>\, .
\end{eqnarray}
Here $<r^2>$ and $<r^4>$ are values of $r^2$ and $r^4$   averaged
over charge density of the nucleus. The low-momentum expansion of
the nuclear electric form factor is of the form
\begin{equation}
\label{FF} F(q^2)\approx
1-\frac{q^2}{6}<r^2>+\frac{q^4}{120}<r^4>\, .
\end{equation}
Modeling the nucleus as a uniformly charged ball  one gets
\begin{equation}
\label{r24}
<r^2>= \frac{3}{5} \ r_0^2, \ \ \ <r^4>= \frac{3}{7} \ r_0^4 \ ,
\end{equation}
where $r_0 =1.1 \ A^{1/3} fm$ is the nucleus radius, and $A$ is
the  nucleus mass number. As one should expect the FNS corrections
(\ref{E0}) obey the following inequalities $\delta
E^{(0)}(2s_{1/2})\gg\delta E^{(0)}(2p_{1/2})\gg\delta
E^{(0)}(2p_{3/2})$.

Let us calculate now the leading in $Z\alpha$  one loop SEVFNS
radiative  correction $E^{(1)}$ for $p_{1/2}$- and
$p_{3/2}$-states. This correction is given by diagrams in
Fig.\ref{Fig1}(b) and Fig.\ref{Fig1}(c). Since we consider the
leading correction, it is sufficient to use the nonrelativistic
approximation for electron wave functions (two-component  wave
functions). It is sufficient also
 to use the effective FNS perturbation that reproduces FNS
correction for s-wave states,
\begin{equation}\label{g}
\delta V_{eff}(r)= g\delta(\bm r)\quad ,\quad g=\frac{2\pi Z\alpha}{3}
<r^2>\, .
\end{equation}
Rest of the calculation is very similar to the  textbook
calculation of the Lamb shift, see, e.g. Ref. \cite{BLP}. We
introduce the parameter $\mu$ such that $m\gg\mu\gg m(Z\alpha)^2$.
Hence the correction $E^{(1)}$ can be represented as a sum of
``high frequency'' and ``low frequency'' contributions
$E^{(1)}=E_h+E_l$, where ``high'' and ``low'' correspond to
frequencies above and below $\mu$, respectively. In the momentum
representation, the effective potential corresponding to the high
frequency contribution is of the form \cite{BLP}
\begin{equation}
\label{fi} \delta \Phi({\bm q})=\left[-\frac{\alpha{\bm q}^2}
{3\pi m^2}\left(\ln\frac{m}{2\mu}+\frac{11}{24}\right)
+\frac{\alpha}{4\pi m}{\bm q} \cdot {\bm \gamma}\right]\delta
V_{eff}({\bm q}) \ ,
\end{equation}
where ${\bm q=\bm p_1-\bm p_2}$ is momentum transfer, and ${\bm
\gamma}$ is the Dirac matrix. Taking the p-wave component of the
potential (\ref{fi}) and transferring it to the coordinate
representation, we get the following expression for the SEVFNS
high frequency correction for a p-wave state $|\psi\rangle$
\begin{eqnarray}\label{Eh}
E_h&=&\frac{g\alpha}{\pi m^2}\langle
\psi|\left\{\frac{2}{3}\left[\ln\frac{m}{2\mu}+\frac{11}{24}\right]({\bm
p}\,  \delta(\bm r)\,{\bm p})
\right.\nonumber\\
&+&\left.\frac{1}{4}(\bm\sigma\bm p)\,
 \delta(\bm r)\,(\bm\sigma\bm p)\right\} |\psi\rangle \ .
\end{eqnarray}
This gives the following values for 2p-states
\begin{eqnarray}\
\label{Ehans}
E_h(2p_{1/2})&=&F\left[\ln\frac{m}{2\mu}+\frac{19}{12}\right]\, , \nonumber\\
E_h(2p_{3/2})&=&F \left[\ln\frac{m}{2\mu}+\frac{11}{24}\right]\, ,\nonumber\\
where \ \  F&=&\frac{\alpha(Z\alpha)^5gm^3}{48\pi^2} \, .
\end{eqnarray}
The contribution of the vacuum polarization, diagram
Fig.\ref{Fig1}(d), can be taken into account in Eqs.
(\ref{Eh}),(\ref{Ehans}) by  substitution $\ln(m/2\mu)
 \to \ln(m/2\mu)-1/5$. The Uehling potential, diagram Fig.\ref{Fig1}(e),
 does not contribute in this order.

The low frequency  contribution is given by the usual
nonrelativistic quantum mechanics expression
\begin{eqnarray}\label{El}
E_l&=&\frac{2g\alpha}{3\pi m^2}\Re \int_0^\mu \omega
d\omega \langle \psi|\,
 {\bm p}\,\frac{1}{E_{2p}^{(0)}-H-\omega+i0}\,
  \delta(\bm r)\, \nonumber\\
&\times&\frac{1}{E_{2p}^{(0)}-H-\omega+i0}
 \,{\bm p}\,|\psi\rangle \, .
\end{eqnarray}
Here $\Re$ stays for real part, $\omega$ is frequency of the
virtual photon, $H=p^2/2m -Z\alpha/r$ is the nonrelativistic
Hamiltonian,  and $E_{2p}^{(0)}=-m(Z\alpha)^2/8$ is the energy of
$2p$-state. We have also taken into account that interaction with
the photon is of the form ${-e\bm p\cdot  \bm A}/m$, where ${\bm
A}$ is the vector potential of the photon. The contribution $E_l$
is the same for $p_{1/2}$- and for $p_{3/2}$-state. Using explicit
form of 2p wave function,
one can represent  (\ref{El}) as
\begin{eqnarray}\label{El1}
E_l&=&\frac{2g\alpha(Z\alpha)^2}{3\pi}\Re\int_0^\mu \omega
d\omega \langle \phi|\,
 \frac{1}{E_{2p}^{(0)}-H-\omega+i0}\,
  \delta(\bm r)\,\nonumber\\
&\times& \frac{1}{E_{2p}^{(0)}-H-\omega+i0}
 \,|\phi\rangle\, ,
\end{eqnarray}
where
\begin{eqnarray}\label{phi}
\phi(r)&=&\frac{1}{\sqrt{32\pi}\,a^{3/2}}(1-r/{6a})\exp{(-r/2a)}\, , \nonumber\\
 a&=&(mZ\alpha)^{-1}\, .
\end{eqnarray}
Eigenvalues of $H$ are $\epsilon_n=-m(Z\alpha)^2/2n^2$. Therefore,
the first impression is that the integrand in Eq. (\ref{El1}) is
singular at $\omega=0$ and $\omega=\epsilon_2-\epsilon_1$.
However, the function $\phi(r)$ is orthogonal to the wave function
$\psi_{2s}(r)$, hence, there is no real singularity at $\omega=0$.
There is a real singularity at $\omega=\epsilon_2-\epsilon_1$ that
is related to the possibility of emission of real photons, and
this slightly complicates integration in (\ref{El1}). To overcome
this technical problem, it is convenient to represent $\phi(r)$ as
$\phi(r)=\varphi(r)+\beta \psi_{1s}(r)$  with
$\beta=\sqrt{2/\pi}(2/3)^4$. In this form $\varphi(r)$ is
orthogonal both to $\psi_{2s}(r)$ and $\psi_{1s}(r)$. Then
(\ref{El1}) is transformed to
\begin{eqnarray}\label{El2}
E_l&=&\frac{2g\alpha(Z\alpha)^2}{3\pi}\Re\int_0^\mu \omega
d\omega \left\{ \left(\langle \bm 0|\,
 \frac{1}{E_{2p}^{(0)}-H-\omega}\,|\varphi\rangle\right)^2 \right.\nonumber\\
&+&\frac{\beta^2\psi_{1s}^2(0)}{(E_{2s}-E_{1s}-\omega+i0)^2}\nonumber\\
&+&\left.\frac{2\beta\psi_{1s}(0)}{(E_{2s}-E_{1s}-\omega+i0)}\,
\langle \bm 0|\,\frac{1}{E_{2p}^{(0)}-H-\omega}\,|\varphi\rangle
 \right\}\, ,
\end{eqnarray}
where $|\bm 0\rangle$ denotes the electron localized at origin. In
this form the matrix element  $\langle \bm
0|\left(E_{2p}^{(0)}-H-\omega\right)^{-1}|\varphi\rangle$ has no
singularities. Using explicit expression for the nonrelativistic
Coulomb Green's function  \cite{Meix}
\begin{eqnarray}\label{gf}
G(0, \bm r \, |E)=\langle \bm r | \frac{1}{E-H}|\bm
0\rangle\nonumber\\
 =\frac{m}{2\pi
r}\Gamma(1-\eta)W_{\eta,1/2}(2pr)\, ,
\end{eqnarray}
where $p=\sqrt{-2mE}$, $\eta= mZ\alpha/p$, $\Gamma$  is the
gamma-function and $W$ is the Whittaker function, and taking the
integral over $r$, and then over $\omega$, we obtain
\begin{eqnarray}\label{Elans}
E_l=F \left[\ln\frac{\mu}{m(Z\alpha)^2}+0.0198\right]\quad .
\end{eqnarray}
Combining (\ref{Ehans}) and (\ref{Elans}), we finally obtain the
total SEVFNS radiative corrections (diagrams  Fig.\ref{Fig1}(b)
and Fig.\ref{Fig1}(c)) in the leading order
\begin{eqnarray}\label{Etot}
&&E^{(1)}(2p_{1/2})=F\left[\ln\frac{1}{(Z\alpha)^2}+0.910\right]\, , \nonumber \\
&&E^{(1)}(2p_{3/2})=F\left[\ln\frac{1}{(Z\alpha)^2}-0.215\right]\, .
\end{eqnarray}
As one should expect, the result is independent of the parameter
$\mu$. We have already mentioned that to account for the  vacuum
polarization (the diagram Fig.\ref{Fig1}(d)) one has to replace
$\ln(1/(Z\alpha)^2) \to \ln(1/(Z\alpha)^2)-1/5$. Therefore the
total FNS radiative corrections (diagrams  Fig.\ref{Fig1}(b),
Fig.\ref{Fig1}(c), and Fig.\ref{Fig1}(d)) in the leading order are
\begin{eqnarray}\label{Etot1}
&&E_{tot}^{(1)}(2p_{1/2})=F\left[\ln\frac{1}{(Z\alpha)^2}+0.710\right]\, ,\nonumber\\
&&E_{tot}^{(1)}(2p_{3/2})=F\left[\ln\frac{1}{(Z\alpha)^2}-0.415\right]\, .
\end{eqnarray}

Let us have a look now at the relative FNS    radiative correction for $2p_{3/2}$ state.
According to Eqs. (\ref{E0}) and (\ref{Etot1}) the relative correction is
\begin{equation}
\label{3/2} \frac{E_{tot}^{(1)}(2p_{3/2})}{\delta
E^{(0)}(2p_{3/2})}= \frac{20}{3\pi}\frac{\alpha <r^2>}{m^2
<r^4>}\left[\ln\frac{1} {(Z\alpha)^2}-0.415\right]\, .
\end{equation}
For example, for Hydrogen atom the radiative correction is by a
factor $2.6 \ 10^4$  larger than the ``leading'' contribution.

According to the present calculation, the leading in powers of
$Z\alpha$ SEVFNS relative radiative correction (diagrams
Fig.\ref{Fig1}(b) and Fig.\ref{Fig1}(c)) for $2p_{1/2}$-state is
equal to (see comment \cite{com1})
\begin{eqnarray}\label{Dp1}
\Delta_p^{(0)}=\frac{E^{(1)}(2p_{1/2})}{\delta E^{(0)}(2p_{1/2})}=
\frac{8\alpha}{9\pi}
\left[\ln\frac{1}{(Z\alpha)^2}+0.910\right]\, .
\end{eqnarray}
As we have already explained, the correction comes from quantum
fluctuations at distances $\lambda_c < r < \lambda_c/(Z\alpha)$.
There is also a contribution
$$
-\frac{\alpha(Z\alpha)^2}{\pi}\left(\frac{15}{4}-\frac{\pi^2}{6}\right)
 \ln(b\lambda_C/r_0)
$$
 that comes from distances $r_0 \ll r \ll \lambda_C$, this contribution
has been calculated in our previous work \cite{Mil}. The
contribution $\propto \alpha(Z\alpha)$ that comes from $r\sim
\lambda_C$ has not been calculated yet. Therefore, altogether one
gets the following formula for  the relative correction
$\Delta_p$:
\begin{eqnarray}
\label{Dp}
\Delta_p&=&-\alpha\left[-\frac{8}{9\pi}
\left(\ln\frac{1}{(Z\alpha)^2}+0.910\right) \right.\\
&+&\left. a_1(Z\alpha)+\frac{(Z\alpha)^2}{\pi}
\left(\frac{15}{4}-\frac{\pi^2}{6}\right)
\ln(b\lambda_C/r_0) \right]\, ,\nonumber
\end{eqnarray}
where $a_1$ is an unknown coefficient. To determine the
coefficient $a_1$, we fit results of numerical calculation of
$\Delta_p$ for $2p_{1/2}$-state \cite{CJS93}. As a result of the
fit we find $a_1=2.75$. The correction $\Delta_p$ given by Eq.
(\ref{Dp}) is plotted in Fig.\ref{Fig2} by the solid line. The
results of computations  \cite{CJS93}  are shown by diamonds.
Agreement is very good.

Concluding, we have shown that corrections to the Lamb shift of
p-wave atomic states due to the finite nuclear size are
qualitatively different from that for s-wave states. The
difference is related to the infrared behavior of quantum
fluctuations. As a result, the leading relative p-wave correction
is proportional to $\alpha\ln(1/Z\alpha)$ while the leading s-wave
correction is proportional to $ \alpha(Z\alpha)$. The leading
p-wave correction has been calculated analytically.

O.P.S. thanks the Institute for Nuclear Theory at the  University
of Washington for its hospitality and the Department of Energy for
partial support during the completion of this work.


\begin{thebibliography}{99}
\bibitem{com} Strictly speaking dependence of $\Delta_s$ on
the principle quantum number
shall appear in higher orders in $Z\alpha$, but this is a very small effect.
\bibitem{com1} Constant in Eq. (\ref{Dp1}) is different from that
presented in the abstract
because  value in the abstract includes the vacuum polarization as well.
\end{thebibliography}
\end{document}